\documentclass[12pt]{article}

\font\twelve=cmbx10 at 15pt
\font\ten=cmbx10 at 12pt

\renewcommand{\thefootnote}{\fnsymbol{footnote}}

\def\beq{\begin{equation}}
\def\eeq{\end{equation}}
\def\bea{\begin{eqnarray}}
\def\eea{\end{eqnarray}}
\def\bq{\begin{quote}}
\def\eq{\end{quote}}

\def\gappeq{\mathrel{\rlap {\raise.5ex\hbox{$>$}}
{\lower.5ex\hbox{$\sim$}}}}

\def\lappeq{\mathrel{\rlap{\raise.5ex\hbox{$<$}}
{\lower.5ex\hbox{$\sim$}}}}

\def\Toprel#1\over#2{\mathrel{\mathop{#2}\limits^{#1}}}

\begin{document}

\begin{titlepage}

\begin{center}

{\ten Centre de Physique Th\'eorique\footnote{UMR 6207 - 
Unité Mixte de Recherche du CNRS et des
Universités Aix-Marseille I, Aix-Marseille II et de l'Université
du Sud Toulon-Var - Laboratoire affilié à la FRUMAM}, CNRS-Luminy, \\
Case 907,  F--13288 Marseille Cedex 9 - France }

\vspace{1.5 cm}

{\twelve 
A SELECTION RULE FOR MULTIQUARK DECAYS }

\vspace{0.3 cm}
\setcounter{footnote}{0}
\renewcommand{\thefootnote}{\arabic{footnote}}

{\bf 
F. Buccella$^1$
}

\bigskip

$^1$ On leave of absence from Dipartimento di Scienze Fisiche,
Universit\`{a} di Napoli ``Federico II", Complesso Universitario di
Monte S. Angelo, Via Cintia, I-80126 Napoli, Italy

\vspace{1.5 cm}

{\bf Abstract}

\end{center}


By assuming $SU(6)_{CS}$ symmetry for pentaquark decays one finds a 
selection rule, which strongly reduces the number of states able to 
decay into a baryon and a meson final state and allows an intriguing
identification for the $\Theta^+$ particle recently discovered  with 
the prediction of a narrow width.

\vspace{\stretch{1}}

\bigskip

\noindent Number of figures : 0
\bigskip

\noindent September 2004
\bigskip

\noindent 
CPT-2004/075 \\
DSF 30/04
\bigskip

\noindent anonymous ftp: ftp.cpt.univ-mrs.fr\\
\noindent www.cpt.univ-mrs.fr

\pagebreak

\end{titlepage}

The discovery of a narrow $Y=2$ $ T=0$ KN resonance $\Theta^+$ at 1540 MeV 
in different experiments \cite{N} is a great success for the Skyrme model 
\cite{S},  which predicts the existence of a $(\overline{10},1/2)^+$ 
state, stable in the non-relativistic limit, in the same group of
states of the better-established $(8,1/2)^+$ and $(10,3/2)^+$
traditionally classified in the 56-dimensional representation of
flavour-spin $SU(6)_{FS}$ \cite{GR}. The value of the mass of
the state happened to be predicted at the right
value \cite{DPP}. In fact, one of the authors (D.D.) has been very
active in promoting the experimental search for that state. 
This state can be thought to be a pentaquark, consisting of four quarks
$uudd$ and a $\bar{s}$.
Pentaquarks have been considered many years ago \cite{HS} and their relevance 
for heavy  quark systems has been stressed \cite{GSRL}.\\
 To get positive parity states with a $S$-wave $\bar{q}$, one
should consider, as in \cite{JW} \cite{SR} \cite{CCKN1} \cite{BGS}  , 
$L = 1$ four-quark states. In a previous paper \cite{BS} all these states 
were
classified in the $126+210+105+105'$ $L=1$ representation of $SU(6)_{FS}
\times SO(3)_L$.\\
Within the approximation of considering the decay of a pentaquark as a
separation process with a $\bar{q}$ forming a meson with one of the
$4q$ -- in the case of $\Theta^+$,  the $\bar{s}$
together with a $u$ or a $d$ forms a $K$ -- and the remaining
three $q$'s giving rise to the final baryon, the pentaquarks with
the $4q$'s transforming as the $105+105'$ representations of $SU(6)_{FS}$
are not allowed to decay into a final state with a meson and a ($8,1/2^+$) 
baryon \cite{SR}. In fact, as it can be easily seen by considering the
Young tableaux associated to those representations, at least two of the 
three quarks remain in a $SU(6)_{FS}$ antisymmetric state, so that
the $3q$'s  wave function is orthogonal to that of the totally symmetric
56 representation. An analogous selection rule has been found
in \cite{BS}, which follows from the fact that the
$(10, \frac{3}{2}^+)$ and the $(8, \frac{1}{2}^+)$ transform as
the 20 and 70 $SU(6)_{CS}$ representations, respectively. The
states of the $210$ and of the $105$ $SU(6)_{CS}$ cannot decay
into meson decuplet states, while the states of the
$\overline{15}$ cannot decay into meson octet states. 
In conclusion, only  the states,
with their $4q$ transforming as the $105'$ of $SU(6)_{CS}$ may be
found by looking for decuplet-meson final states in octet-meson reactions.\\
$SU(6)_{CS}$ plays an important role in the mass splittings of the
$L=0$ ordinary hadrons ($3q$ baryons and $q\bar{q}$ mesons) and 
the chromomagnetic interaction, which predicts properly \cite{DGG}
the $\Delta-N$ and $\rho-\pi$ mass splittings, gives a contribution 
proportional to a combination of the $SU(6)_{CS}$, $SU(3)_C$ and $SU(2)_S$ 
Casimir operators \cite{HS}. This fact was the motivation \cite{BS} to 
write a mass  formula for pentaquarks, where the mass splittings are 
provided by the chromomagnetic interaction of the $4q$'s and the $\bar{q}$ 
and by a spin-orbit term, as for ordinary hadrons.

In \cite{BS} the phenomenogical mass formula has been proposed:
\bea
m&=&m_0+h\frac{3}{16}( {m_{K^*}-m_K} )
\left[ C_6(p)-C_6(t)-\frac{1}{3}S_p(S_p+1)+\frac{1}{3}
S_t(S_t+1) \right. \nonumber \\
&& \left. -\frac{4}{3} \right] + 
\tilde{h}\frac{1}{4}( m_N - m_{\Delta} ) \left[ C_6(t)-\frac{1}{3}
S_t(S_t+1)-\frac{26}{3} \right] +a \,\vec{L} \cdot \vec{S}
\label{equation1}
\eea
where $C_6(p)$ and $C_6(t)$ are  the Casimir of $SU(6)_{CS}$ and
$p$ and $t$ are the representations for pentaquark and $4q$ states,
respectively. In \cite{BS} the values $h=1/2$, $\tilde{h}=0$
and $a=40MeV$ have been chosen for the positive parity states
built with ($4q,L=1$) and a $\bar{q}$ in $S$-wave respect to them,
$h=\tilde{h}=1$ for the negative parity states with all the constituent
in $S$-wave and $h=0$, $\tilde{h}=1$ for the states built with $4q,L=0$
and a $\bar{q}$ in $P$-wave respect to them; in this case, inspired by 
the spectrum of the mesons of the ($35,L=1$) of $SU(6)_{FS}$, we take
$a=100MeV$ \cite{B}.\\  
That choice implies that the transformation properties with respect to
$SU(6)_{CS}$ of the $L=1$ $4q$'s and of the pentaquarks, one forms
combining them with the $\bar{q}$, play a major role to identify
the mass eigenstates. In Table 1 we write for the ($4q,L=1$)  
$SU(3)_F \times SU(2)_2$ multiplets  the transformation properties
with respect to $SU(6)_{CS}$ and $SU(6)_{FS}$.     

\begin{center}
\begin{tabular}{|c|c|c|}\hline
$SU(6)_{CS}$ & $SU(6)_{FS}$ & $SU(3)_F \times SU(2)_S$ 
\\
\hline

105'        &    126            &      $(15', S= 2)$ \\
105'        &    210            &      $(15, S=2 )+(15', S = 1)$ \\      
105'        &    105            &      $(\bar{6}, S = 2)+(15', S = 0)$ \\
105'        &    105'           &      $(3, S = 2)$ \\
\hline
 & & \\
105'+210    &    126+105        &      $(\bar{6}, S = 0)$ \\    
105'+210    &    210+105'       &      $(\bar{6}, S = 1)+(15+3, S = 0)$ \\
\hline
 & & \\
105'+210+105 &  210+105+105'   &      $(3 , S = 1)$ \\ 
\hline
 & & \\
105'+210+105+$\bar{15}$  &  126+210+105+105'     &  $(15, S = 1)$ \\ 
\hline
\end{tabular}
\end{center}
\begin{center}
TABLE 1 : Trasformation properties with respect to $SU(6)_{CS}$,
$SU(6)_{FS}$ and $SU(3)_F \times SU(2)_S$ of the ($4q,L = 1$)
states. For convenience, the $SU(2)$ representations are not denoted
by their dimensions - which is the case for their $SU(3)$ partners
- but by their highest weight. Apart the states, which transform as a
$15'$ of $SU(3)_F$ or (and) a $S = 2$ of $SU(2)_S$, the states with 
definite  transformation properties with respect to $SU(6)_{CS}$ are a 
combination of states with definite $SU(6)_{FS}$ properties.In the case 
of  the $\bar{6}$  with $S = 1$ and $0$, the mixing is maximal.
\end{center} 
In Tables 2 and 3 we report the mass splittings for the ($Y=2$,$I=0$) 
pentaquark states deduced from eq.(1) and the values chosen for the 
parameters.\\

\begin{center}
\begin{tabular}{|c|c|c|c|}\hline
$SU(6)_{CS} \times S$ & $J$ & $\Delta M$ ($MeV$) \\
\hline (20*, $\frac{3}{2}$)(105') &  $\frac{5}{2}$ + $\frac{3}{2}$ +
$\frac{1}{2}$ & -150 + 40$(\vec{L}
\cdot \vec{S})$ \\
\hline (70*, $\frac{1}{2}$)(210) & $\frac{3}{2}$ + $\frac{1}{2}$ 
& -190 + 40$(\vec{L} \cdot \vec{S})$ \\
\hline (70*, $\frac{1}{2}$)(105') & $\frac{3}{2}$ + $\frac{1}{2}$ 
& -48 + 40$(\vec{L} \cdot \vec{S})$\\
\hline (540, $\frac{5}{2})$(105') & $\frac{7}{2}$ + $\frac{5}{2}$ +
$\frac{3}{2}$ & +64 + 40$(\vec{L}
\cdot \vec{S})$\\
\hline (1134, $\frac{3}{2}$)(210) & $\frac{5}{2}$ + $\frac{3}{2}$ + 
$\frac{1}{2}$ & +76 + 40$(\vec{L}
\cdot \vec{S})$ \\
\hline (540*, $\frac{3}{2}$)(105') & $\frac{5}{2}$ + $\frac{3}{2}$ +
$\frac{1}{2}$ & +92 + 40$(\vec{L}
\cdot \vec{S})$\\
\hline (1134*, $\frac{1}{2}$)(210) & $\frac{3}{2}$ + $\frac{1}{2}$ 
& +95 + 40$(\vec{L} \cdot \vec{S})$ \\
\hline (540*, $\frac{1}{2}$)(105') & $\frac{3}{2}$ + $\frac{1}{2}$ 
& +95 + 40$(\vec{L} \cdot \vec{S})$\\
\hline (70, $\frac{1}{2}$)(105)     & $\frac{3}{2}$ + $\frac{1}{2}$ 
& -98 + 100 $\vec{L} \cdot \vec{S}$\\
\hline (560, $\frac{3}{2}$)(105)   & $\frac{5}{2}$ + $\frac{3}{2}$
+ $\frac{1}{2}$  &  -98 + 100 $\vec{L} \cdot \vec{S}$\\
\hline
\end{tabular}
\end{center}
\vspace*{0.1cm}
\begin{center}
Table 2 ©: Mass splittings of the positive parity ($Y=2,I=0$) 
pentaquarks.\\
The first eight rows correspond to the states with ($4q,L=1$) and a 
$\bar{q}$ in $S$-wave. The last two rows to ($4q,L=0$) and a $\bar{q}$
in $P$-wave.
The * is put to remind of a mixing between the $ SU(6)_{CS}$
representations and the transformation properties of the $4q$
state have been written in brackets.
\end{center}
\vspace*{0.1cm}
\begin{center}
\begin{tabular}{|c|c|c|}\hline
$SU(6)_{CS}$             & $J=S$                     &   $\Delta M$ ($MeV$)\\

\hline 70(105)           & ($\frac{1}{2}$ )          & $-342$   \\

\hline 560(105)          & ($\frac{3}{2}$)           & $+24$  \\

\hline
\end{tabular}
\end{center}
\begin{center}
Table 3 : Mass splittings of the (Y=2,I=0) negative parity pentaquarks 
built with $4q$ and a $\bar{q}$ in S-wave.
\end{center}
\vspace*{0,1cm}

The dynamics for pentaquark decays may be different from the case of
$\Delta$ and $\rho$ decays, where one has to create a $q\bar{q}$ pair,
with the $\bar{q}$ forming a meson with one of the initial quarks in the
first case or with the initial quark in the second one.
As long as for pentaquarks all the elementary fermions in the final state
are present in the initial one, which makes possible the hypothesis that
the decay is a consequence of the separation of its constituents. Also, 
at difference with what happens for the decay of the previously mentioned 
ordinary  hadrons,with the orbital angular momentum not conserved (changing
from 0 to 1) as well as the spin, for the pentaquark decay, to the initial
orbital momentum of the $4q$'s in the initial state corresponds the 
relative angular momentum of the emitted meson with respect to the final
baryon.
So $L$ and $S$ may be both conserved. One may even assume that, as in the 
hypothesis that the amplitude is proportional to the scalar product
of the initial and final wave-functions, also $SU(6)_{CS}$ and (or)
$SU(6)_{FS}$ are conserved in pentaquark decays.
We want first to explore the consequences of $SU(6)_{CS}$ conservation for
the decay of a pentaquark into a meson baryon final states, which are very
restrictive, since the pseudoscalar mesons are $SU(6)_{CS}$ singlets.
According to the previously mentioned transformation properties of the
baryon of the 56 of $SU(6)_{FS}$, only the pentaquarks transforming as a 
70 (20) of $SU(6)_{CS}$ may decay into a final state containing a pseudoscalar
meson and a $(8,1/2)^+$ ($(10,3/2)^+$) baryon. As long as the 
$\overline{10}$ multiplets of $SU(3)_F$, they may be obtained by combining the
$\bar{6}$ constructed with $4q$ with the $\bar{q}$, which transforms as a 
$\bar{3}$ of $SU(3)_F$. According to Table1 only the 105' and the 210 
representations of $SU(6)_{CS}$ contain $\bar{6}$'s of $SU(3)_F$ (the demand 
of complete antisymmetry for the $L=1$ $4q$ wavefunction relates the 
transformation properties with respect to different groups \cite{BS}). Let us 
therefore consider the $SU(6)_{CS}$ products:

\bea
210 \times \bar{6} &=& 1134 + 56 +70 
\label{equation2}
\eea
\bea
105' \times \bar{6}&=& 540 + 70 +20
\label{equation3}
\eea

In conclusion the only $\overline{10}$ states allowed to decay into a 
pseudoscalar meson and a $(8,1/2)^+$ baryon should transform as a 70 of 
$SU(6)_{CS}$, These are the combinations written in \cite{BS} gor the
$4q,L=1$ states:
\newpage
\bea
&|70, & (1,S = 1/2), S_z = \frac{1}{2} > = \nonumber \\
& \frac{1}{\sqrt{3}} &  \{ \frac{1}{\sqrt{3}} \; | 105' (3, S = 1)
S_z = 1>_a \;
|\bar{6} ; \, (\bar{3}, S= 1/2), S_z = - 1/2>^a \nonumber \\
& \frac{-1}{\sqrt{6}} & | 105' (3, S = 1) S_z = 0>_a \;
|\bar{6} ; \, (\bar{3}, S = 1/2), S_z =  1/2>^a \} \nonumber \\
& + \frac{1}{\sqrt{2}} & | 105' (3, S = 0) >_a \;
|\bar{6} ; \, (\bar{3}, S = 1/2), S_z =  1/2>^a \} \nonumber \\
\label{equation4}
\eea

\bea
&|70, & (1,S = 1/2), S_z = \frac{1}{2} > = \nonumber \\
& \frac{1}{\sqrt{3}} &  \{ \frac{1}{\sqrt{2}} \; | 210 (3, S = 1) S_z
= 1>_a \;
|\bar{6} ; \, (\bar{3}, S= 1/2), S_z = - 1/2>^a \nonumber \\
& -\frac{1}{2} & | 210 (3, S = 1) S_z = 0>_a \;
|\bar{6} ; \, (\bar{3}, S = 1/2), S_z =  1/2>^a \nonumber \\
& +\frac{1}{2} & | 210 (3, S = 0) >_a \;
|\bar{6} ; \, (\bar{3}, S = 1/2), S_z =  1/2>^a \} \nonumber \\
\label{equation5}
\eea

where $a = 1,2,3$ is a colour index to be saturated to get a colour singlet, 
and for the $4q,L=0$ state: 

\bea
&|70, & (1,S = 1/2), S_z = \frac{1}{2} > = \nonumber \\
& \frac{1}{\sqrt{3}} &  \{ \sqrt{\frac{2}{3}} \; | 105 (3, S = 1)
S_z = 1>_a \;
|\bar{6} ; \, (\bar{3}, S= 1/2), S_z = - 1/2>^a \nonumber \\
& \frac{-1}{\sqrt{3}} & | 105' (3, S = 1) S_z = 0>_a \;
|\bar{6} ; \, (\bar{3}, S = 1/2), S_z =  1/2>^a \nonumber \\
& +\frac{1}{2} & | 210 (3, S = 0) >_a \;
\label{equation6}
\eea
for the $4q,L=0$ states.
The mass eigenstates are approximately given by states with definite 
$SU(6)_{CS}$ and the two $(\overline{10},1/2)^+$ states, built with
$4q,L=1$ and a $\bar{q}$ in $S$-wave respect to them, have their larger 
components along the $70$'s of $SU(6)_{CS}$ and masses, which differ by about
$140MeV$. By identifying the lightest one with the  discovered  $\Theta^+$ 
state, one predicts the existence of another ($\overline{10},1/2^+$) 
resonance at about $1680MeV$.\\
As stated in the caption of Table1 the ($\bar{6},S=1$) and ($\bar{6},S=0$) 
states of the $210$ representation of $SU(6)_{CS}$ are maximal mixtures of 
the $210$ and $105'$ and $126$ and $105$ representations of $SU(6)_{FS}$, 
respectively. The $SU(6)_{FS}$ selection rule implies that only the first 
components, the $210$ and the $126$ of $SU(6)_{FS}$ contribute to the decay 
into a final state consisting of a pseudoscalar meson and of a $(8,1/2)^+$ 
baryon (as $KN$). In \cite{CCKN2} the amplitude for the decay $\Theta^+$ into  
$KN$ is predicted to be suppressed by the small overlap factor $\frac{5}{96}$, 
if one assumes that the $4q$'s in the $\Theta^+$ transform as the 
($126, \bar{6}, S=0, L=1$) of  $SU(6)_{FS} \times SU(3)_F \times SU(2)_S 
\times SO(3)_L$.\\ 
Our $SU(6)_{CS}$ selection rules implies that the $ S = 1/2 $ colour  singlet 
of the  $1134$ representation of $ SU_{CS}(6)$ cannot decay into a pseudoscalar
$1/2$ octet final state. From the ortogonality of the $70$ and $1134$
final states and the vanishing coupling to the $KN$ final state of the 
second one, one can relate  the couplings of the $S=1$ and $S=0$ states and, 
consequently, the coupling of the state in the left hand side of  eq.(5) 
to the one of the ($\overline{10}, J=1/2$) state constructed with  the
tetraquark in the ($126, \bar{6}, S=0, L=1$)  of  $SU(6)_{FS}$ and the
$\bar{q}$: in fact the ratio of the two couplings  comes out to be 
$\sqrt{2}$  and so, by identifying the $\Theta^+$ with the l.h.s of  eq.(5), 
we predict a width twice larger than in \cite{CCKN2} with a different 
state with the diagonalization fixed by the mass formula in Eq.(1), where
the chromomagnetic interaction, successfully introduced  in \cite{DGG}
for ordinary hadrons, plays the main role.\\
By considering ($4q$,$L=0$), we can build positive (negative) parity 
states by combining them with a $\bar{q}$ with $L=1(0)$ with respect to 
them. In both cases the $4q$ $\bar{6}$ of $SU_F(3)$ should transform as
the $105$ representation of $SU_{CS}(6)$ with $S=1$.
By combining it with the $\bar{6}$ of $SU_{CS}(6)$:
\begin{equation}
126 \times \bar{6} = 560 + 70
\label{equation7}
\end{equation}   
one realizes that only the $\overline{10}$'s with $S=1/2$, which transform 
as the $70$ of $SU(6)_{CS}$, are allowed to decay into a pseudoscalar $1/2$ 
octet state.\\
Within the approximation first suggested in \cite{HS} of requiring that
the $\bar{q}$ should form a meson only with a $q$ in the some cluster,
one expects a narrow width for the $J=1/2^+$ state built with ($uudd,I=0,S=1,L=0$) and a $\bar{s}$ in 
$P$-wave respect to them, transforming as the $70$ representation of 
$SU(6)_{CS}$, which according to Table 2 is $32MeV$ heavier than the 
state, we have identified with the $\Theta^+$ positive parity. The 
$J=1/2^+$ state transforming as the $560$ representation of $SU(6)_{CS}$
is even lighter, but is forbidden also by the $SU(6)_{CS}$ selection rule
to decay into a $KN$ final state.\\ 
One may write for the positive parity ($qq\bar{q} \bar{q},L=0$) ($q=u,d$) 
meson states a mass formula analogous to eq.(1):
\bea
\mu&=&\mu_0+\frac{3}{16} ( m_{\rho}- m_{\pi} )
\left[C_6(\tau)-C_6(2q)-C_6(2\bar{q})+C_3(2q)
-\frac{1}{3}C_2({\tau}) \right. \nonumber \\
&& \left. + \frac{1}{3}C_2(2q)) 
 +\frac{1}{3}C_2(2\bar{q}) \right] + 
\frac{1}{4}( m_N - m_{\Delta} )
\left[ C_6(2q)+C_6(2©\bar{q})-C_3(2q) \right. \nonumber \\
&& \left. - \frac{1}{3}C_2({2q})-\frac{1}{3} C_2(2\bar{q})-8 \right]
\label{equation8}
\eea
where $\tau$ is the $2q2\bar{q}$ state.

The lightest state, with a contribution of the chromo-magnetic interaction 
$\simeq - 1 GeV$, is a ($ I = 0 , 0^+ $) state with quark content 
$ud\bar{u}\bar{d}$ \cite{J}, which transforms mostly as a singlet of 
$SU(6)_{CS}$, to  be identified with the $f^0(600)$ $0^+$ state \cite{J}. 
With the  appropriate  changes for the presence of strange quarks several 
hundreds  $MeV$ above that state one predicts a ($I = 0 + 1, 0^+$)  
$qs\bar{q}\bar{s}$ multiplet to be identified with the 
($f^0(980)+a^0(980),0^+$) states, for which the $qs\bar{q}\bar{s}$ content 
has been already proposed \cite{WI}.\\
According to $SU(6)_{CS}$ symmetry only $SU(6)_{CS}$ singlets, as the 
states just mentioned, may decay into two pseudoscalar mesons.
 For the same reason only the $qq\bar{q}\bar{q}$ 
states transforming as the $35$ representation of $SU(6)_{CS}$ should
be allowed to decay into a final state consisting of a pseudoscalar and
a vector meson.\\
The OZI selection rule \cite{OZI} forbids the decay 
$f^0(980)\rightarrow \pi \pi$ and accounts for the relevance 
in that region, despite the larger phase space for $\pi\pi$, 
of the $K \bar{K}$ channel, which plays an important role in the
disprove \cite{ACCGL} of the lower bound found \cite{Y} for the pion radius.\\
The symmetry with respect to $SU(6)_{FS}$ would also have important
consequences. In fact by eqs.(2-4) and the tensor products:
\begin{equation}
56 \times 35 = 1134 + 700 + 70 + 56
\label{equation9}
\end{equation}
\begin{equation}
105 \times \bar{6} = 560 + 70
\label{equation10}
\end{equation}   
we reach the conclusion that the pentaquarks transforming as the
exotic $SU(3)_F$ representations cannot decay into the final
state consisting of a pseudoscalar or a vector meson  and a
baryon of the octet $1/2©^+$ or of the decuplet $3/2^+$, if
their $4q$'s transform as the $105 + 105'$ of $SU(6)_{FS}$
\cite{SR}, have their couplings to these states proportional
with ratios dictated by $SU(6)_{FS}$ symmetry if their $4q$'s
transform as the $126$ or $210$.\\

\end{document}